\documentclass[]{elsart3p}

\usepackage{graphicx}

\usepackage{amssymb}

\usepackage{lineno}

\begin{document}

\begin{frontmatter}



\title{Change of strength of vortex pinning in YBCO due to BaZrO$_3$ inclusions}


\author[RomaTre]{N. Pompeo\corauthref{cor}},
\corauth[cor]{Corresponding author.}
\ead{\\pompeo@fis.uniroma3.it}
\author[ENEA]{V. Galluzzi},
\author[RomaTre]{R. Rogai},
\author[ENEA]{G. Celentano},
\author[RomaTre]{E. Silva}

\address[RomaTre]{Dipartimento di Fisica ``E. Amaldi'' and Unit\`a CNISM,
Universit\`a Roma Tre, Via della Vasca Navale 84, 00146 Roma,
Italy}

\address[ENEA]{ENEA-Frascati, Via Enrico Fermi 45, 00044 Frascati, Roma, Italy}

\begin{abstract}
We probe the short-range pinning properties with the application of microwave currents at very high driving frequencies (47.7 GHz) on YBa$_2$Cu$_3$O$_{7-\delta}$ films with and without sub-micrometer BaZrO$_3$ inclusions. We explore the temperature and field ranges 60 K$<T<T_c$ and 0$<\mu_0H<$0.8 T, with the field applied along the c-axis. The magnetic field induces a much smaller increase of the microwave resistivity, $\Delta \rho_1(H)+\mathrm{i}\Delta \rho_2(H)$, in YBa$_2$Cu$_3$O$_{7-\delta}$/BaZrO$_3$ with respect to pure YBa$_2$Cu$_3$O$_{7-\delta}$. $\Delta \rho_1(H)$ is slightly superlinear in pure YBa$_2$Cu$_3$O$_{7-\delta}$ (suggesting a possible contribution of thermal activation), but linear or sublinear in YBa$_2$Cu$_3$O$_{7-\delta}$/BaZrO$_3$ (suggesting a possible suppression of thermal activation as a consequence of BaZrO$_3$ inclusions). These features persist up to close to $T_c$. We discuss our data in terms of the ratio $r=\Delta X_s'(H)/\Delta R_s'(H)$ in the framework of the models for the microwave surface impedance in the mixed state. Large $r$ are found in YBa$_2$Cu$_3$O$_{7-\delta}$/BaZrO$_3$, with little field dependence. By contrast, smaller values and stronger field dependences are found in pure YBa$_2$Cu$_3$O$_{7-\delta}$. We discuss the different field dependence of the pinning constant.
\end{abstract}

\end{frontmatter}

\section{Introduction}
\label{intro}

The dynamic of flux lines has shown a surprising richness in high-$T_c$ compounds \cite{blatterRMP94}, due to the several energy scales involved, whose competition is enhanced by high operating temperatures, anisotropy, short coherence length and high $\kappa$. In particular, it was early recognized that  in a very wide portion of the $H-T$ phase diagram flux lines were able to move, and very difficult to pin. Columnar defects \cite{civalePRL91} produced by heavy-ion irradiation have been shown to strongly pin flux lines, thanks to the much increased correlation of the fluxons along the columns. By contrast, point defects change the nature of the vortex transition in the direction of a more disordered (glassy) state \cite{fendrichPRL95}. Recently, a new class of pinning centers has been the subject of strong interest. In fact, it has been shown that BaZrO$_3$ (BZO) inclusions substantially increased critical currents and irreversibility fields in YBa$_2$Cu$_3$O$_{7-\delta}$ (YBCO) films. In particular, it was found that also in strongly pinned YBCO films the critical current density could be increased by a large factor even in fields of several tesla  \cite{macmanusNATMAT04}. Moreover, depending on the size of BZO crystallites, the irreversibility field was raised above 10 T at 77 K \cite{peurlaPRB07}.\\
Apart the obvious practical interest for applications, the features induced by BZO inclusions in the vortex matter are of interest. It was shown by angular measurements \cite{macmanusNATMAT04} that such crystallites induce a columnar-like flux pinning, with a maximum of $j_c$ along the c-axis. It is then interesting to investigate the behaviour of the vortex system in YBCO/BZO materials also by different techniques.\\
Up to now, the vortex properties in YBCO/BZO have been evaluated by means of dc probes: resistivity, critical current, irreversibility line. In all these cases one probes large motion of the flux lines: significant finite resistivity and critical currents are a consequence of large mean values of the vortex displacement. In this regime the vortex-vortex interactions may be very significant, and stiffness of the vortex lattice plays a major role. With increasing driving frequency, vortices are forced to oscillate for very small amplitudes around their equilibrium position. At microwave ($\omega/(2\pi)$=1-100 GHz) frequencies, the mean vortex displacement can be as small as a fractions of nm \cite{tomaschPRB88}. Eventually, with increasing frequency the response is dictated by single-vortex properties: pinning strength, given by the shape and the height of the pinning potential, line tension, and in general the vortex structure (if a rigid or flexible rod, a correlated or not correlated stack of pancake vortices, etc.). An example of this peculiar feature of the high frequency probe is the measurement of the surface impedance in YBCO crystals in the several tens of GHz range \cite{tsuchiyaPRB01}, where absolutely no signature of the vortex melting was found. By contrast, in Bi$_2$Sr$_2$CaCu$_2$O$_{8+x}$ \cite{hanaguriPRL99} clear effects of the first order vortex transition on the surface impedance were observed. This is understandable within the single-vortex response: in YBCO across the transition the vortices retained their 3D, well correlated nature, while in Bi$_2$Sr$_2$CaCu$_2$O$_{8+x}$ correlation across the $c$ axis was most probably lost. Similarly, the implantation of columnar defects gave small but measurable differences in YBCO at 50 GHz \cite{silvaIJMPB00}. In short, a change in the vortex structure or in the interaction with pinning centers gives rise to a difference in the microwave response.\\
The dynamics of flux lines at microwave frequencies has been studied by several authors \cite{gr,cc,brandtPRL91,kv}. Here we will use the model developed by Brandt \cite{brandtPRL91}, where the role of the activation energy is clear. There, the microwave fluxon resistivity can be written as:
\vspace{-1mm}
\begin{equation}
\label{eqB}
    \Delta\rho_1+\mathrm{i}\Delta\rho_2=
    \rho_{ff}\frac{1+\mathrm{i} \omega\tau_{r}}{1/\epsilon'+\mathrm{i} \omega\tau_{r}}\rightarrow\rho_{ff}\frac{1+ \mathrm{i} \frac{\omega_p}{\omega}}{1+\left(\frac{\omega_p}{\omega}\right)^2}
\end{equation}
\vspace{-1mm}
where $\rho_{ff}$ is the flux flow resistivity, $\tau_r=\tau_0\mathrm{e}^{U/k_BT}$ is the characteristic time for thermally activated depinning (creep), $\tau_0=\eta/k_p$ is the relaxation time for elastically pinned vortices, $k_p$ is the pinning constant, $\eta$ is the vortex viscosity, $U$ is a vortex activation energy and $\omega_p=k_p/\eta$ is the depinning angular frequency. Finally, $\epsilon'=1/(1+\mathrm{e}^{U/k_BT})\geq0$ is a dimensionless creep parameter. The latter relation in (\ref{eqB}) corresponds to the Gittleman-Rosenblum (GR) model \cite{gr}, and it holds when thermal activation can be neglected  ($U, \tau_r\rightarrow\infty$). Whichever the model chosen, it is often particularly useful to focus on the so-called $r$ parameter \cite{halbritter}, as defined by $r=\Delta\rho_2(H)/\Delta\rho_1(H)$, because this is in most cases an experimental quantity that can be obtained without the introduction of more or less invasive calibrations of the data. On the other hand, if one takes advantage of a specific model, the physical meaning of $r$ further increases. In particular, focusing on pinning-related processes it is easy to show that, according to Eq. (\ref{eqB}), $r$ gives a lower bound for the normalized pinning frequency $\omega_p/\omega$ and allows to estimate the maximum value of the creep parameter $\epsilon'_{max}=1+2r^2-2r\sqrt{1+r^2}$.

\section{Experimental results and discussion}
\label{exp}
\begin{figure}
\begin{center}
\includegraphics[width=6.2cm]{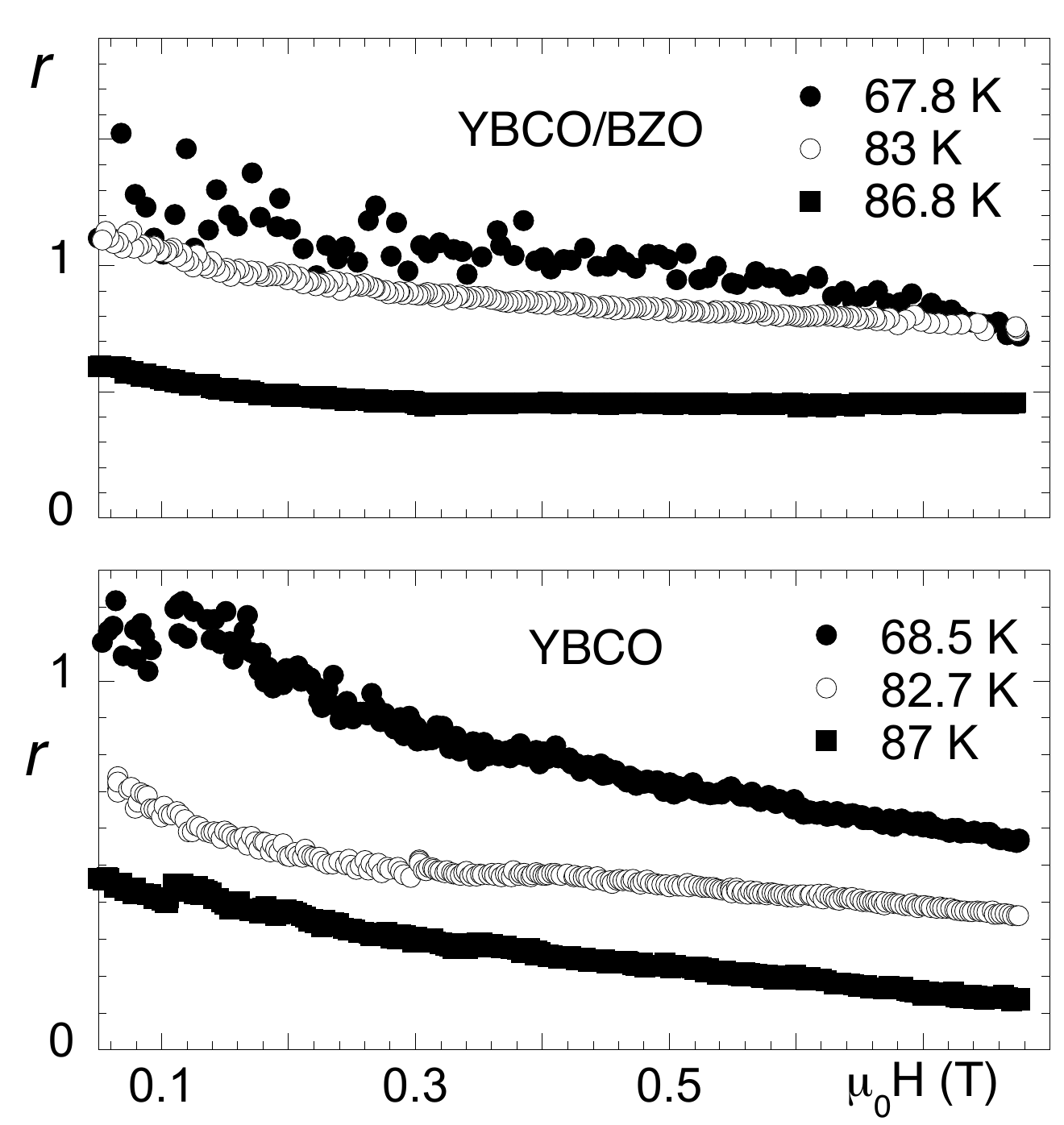}
\end{center}
\vspace{-4mm}
\caption{Field variation of the $r$ parameter in YBCO/BZO and YBCO. Note the different vertical scales.}
\label{fig_r}
\end{figure}

YBCO films were grown by high-oxygen pressure pulsed laser deposition technique on (001) SrTiO$_3$ (STO) substrate. YBCO targets pure and added with 7 mol.\% BZO powder with granularity below 1 $\mu$m were used to grow the films \cite{galluzziIEEE07}. In films with BZO concentration similar to the one here studied the defect concentration, evaluated following \cite{damAPL94}, was larger by a factor $\sim$ 30. The YBCO/BZO films exhibited only moderate changes of the transport properties in zero field, in dc as well as at microwave frequencies. Dc resistivity yielded $T_{c0}=90.2$ K and 89.9 K with $\Delta T_{c}(10\%-90\%)$=0.8 K and 1.2 K in YBCO and YBCO/BZO, respectively. \\
A sapphire dielectric resonator technique was used to get the filed-dependent changes in the complex microwave response at 47.7 GHz \cite{pompeoJSUP07}. We verified that measurements were taken in the linear response regime. Care was taken to avoid the substrate resonances introduced by the strong temperature dependence of the permittivity of the substrate \cite{pompeoPREP07}. From the resonator parameters we obtained the microwave resistivity $(\Delta \rho_{1}+\mathrm{i}\Delta \rho_2)/d$, with the film thickness $d\sim$ 120 nm.\\
\begin{figure}
\begin{center}
\includegraphics[width=6cm]{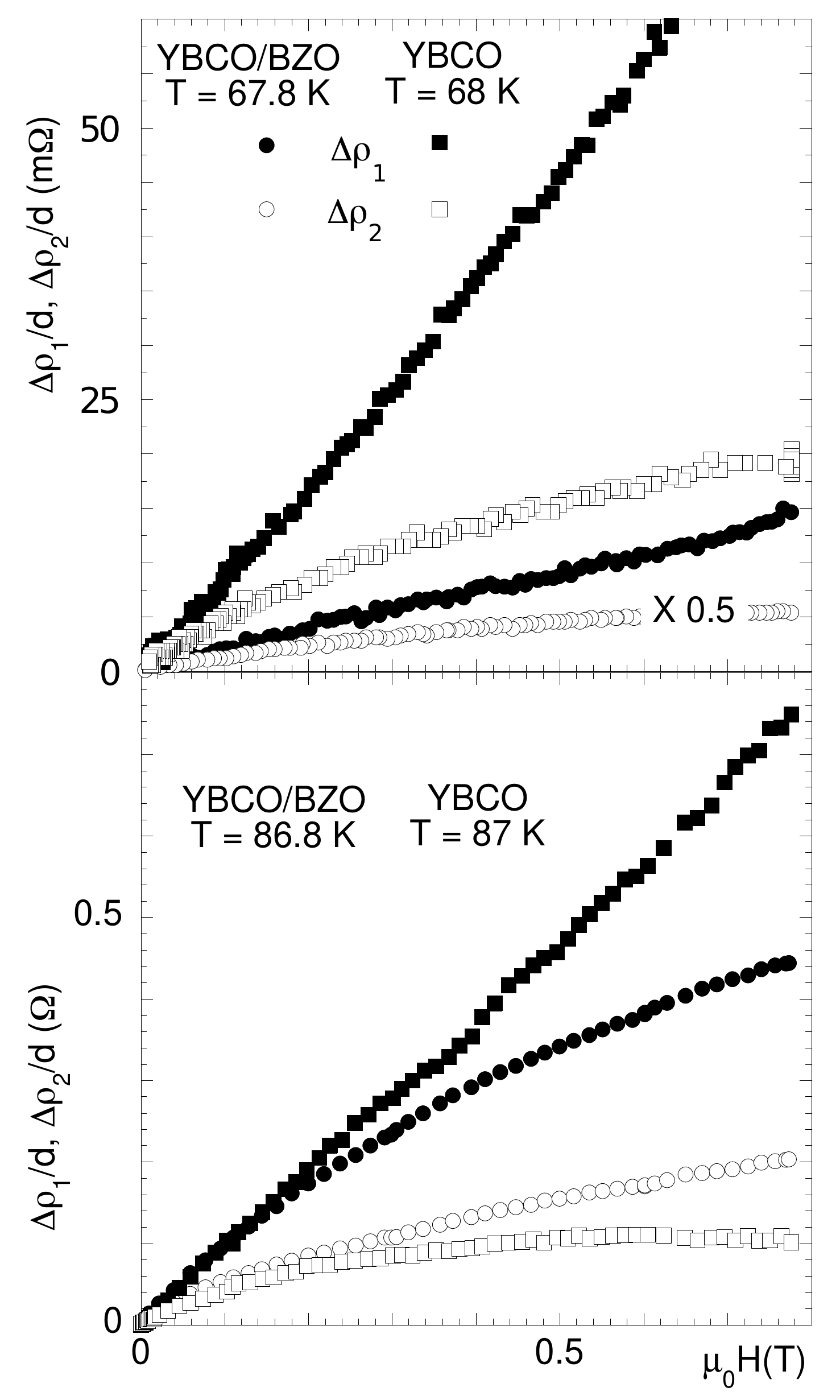}
\end{center}
\vspace{-4mm}
\caption{Field variation of the complex resistivity in pure YBCO (squares) and in YBCO/BZO (circles) at two temperatures.}
\label{fig_rho}
\end{figure}
Typical data of the microwave complex resistivity are reported in Figure \ref{fig_rho} for pure YBCO and YBCO/BZO. The response of the two samples is strikingly different: in YBCO/BZO the same field, at the same temperature, induces a resistivity smaller by a factor $\sim$3 than in pure YBCO (this effect is unlikely to arise from increased sample inhomogeneities: in that case, a larger magnetoresistance would be expected). Remarkably, a significant reduction of the real part persists up to high temperatures. Moreover, the imaginary part is of the same order than the real part in YBCO/BZO also at high $T$, indicating rather strong pinning. We recall that the reduction of the microwave resistivity at similar frequencies due to columnar defects was only $\sim$ 15\% \cite{civalePRL91}.\\
We now focus on the field dependence of the $r$ parameter. In Figure \ref{fig_r} we report $r(H)$ at several temperatures (the numerical derivation of $r$ yields large uncertainty in the low field region, which is omitted). Again, it is readily seen the different behaviour of YBCO and YBCO/BZO. First, the absolute values of $r$ are larger in YBCO/BZO, indicating a stronger pinning. Second, the field dependence is weaker in YBCO/BZO. This is particularly significant at high temperatures: even at 87 K, close to $T_c$, $r$ saturates in YBCO/BZO at $\sim$0.5, while in pure YBCO $r$ keeps decreasing below $\sim$0.2.

\begin{figure}[h]
\centerline{\includegraphics[width=8.3cm]{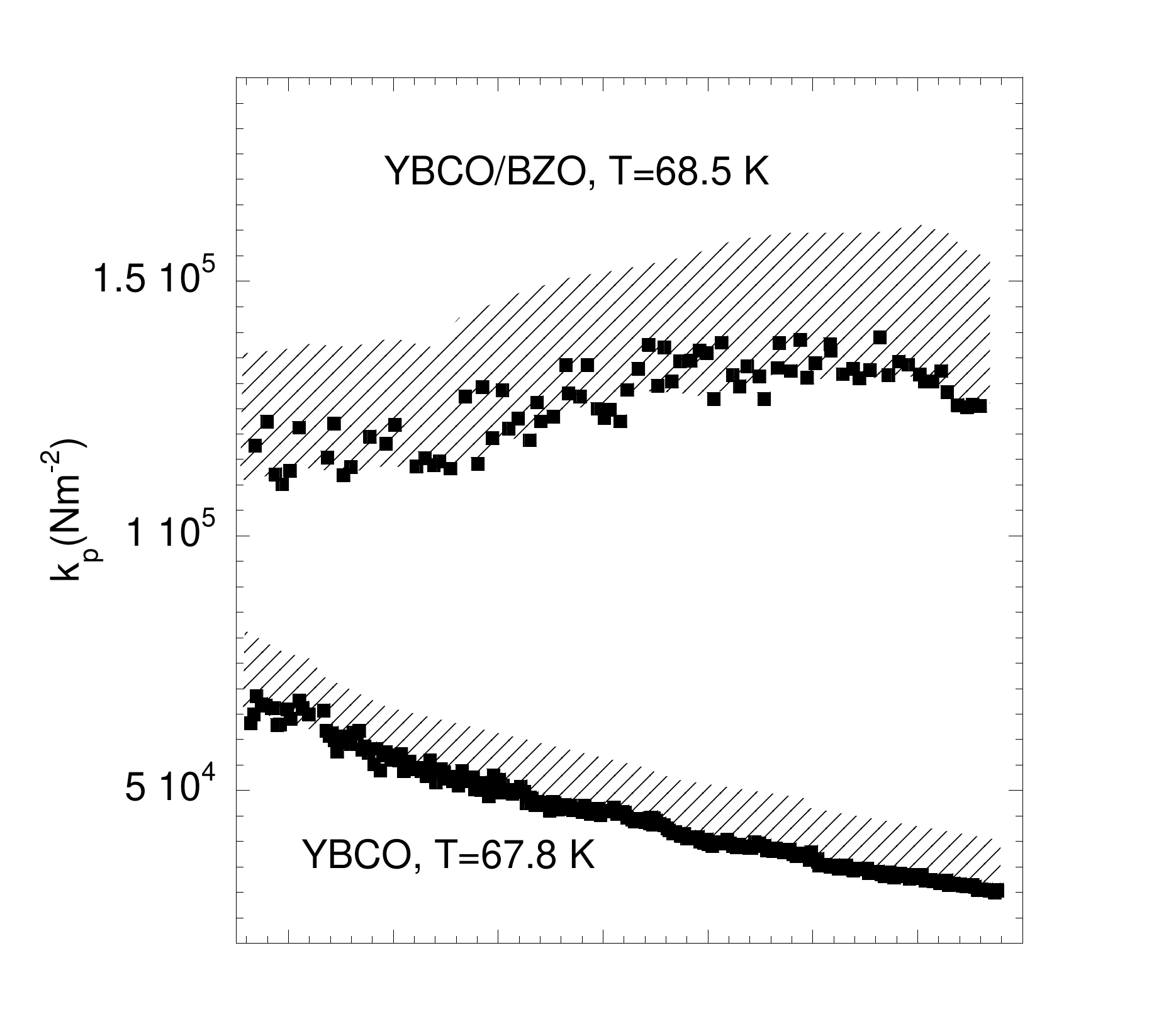}}
\vspace{-9mm}
\centerline{\includegraphics[width=8.3cm]{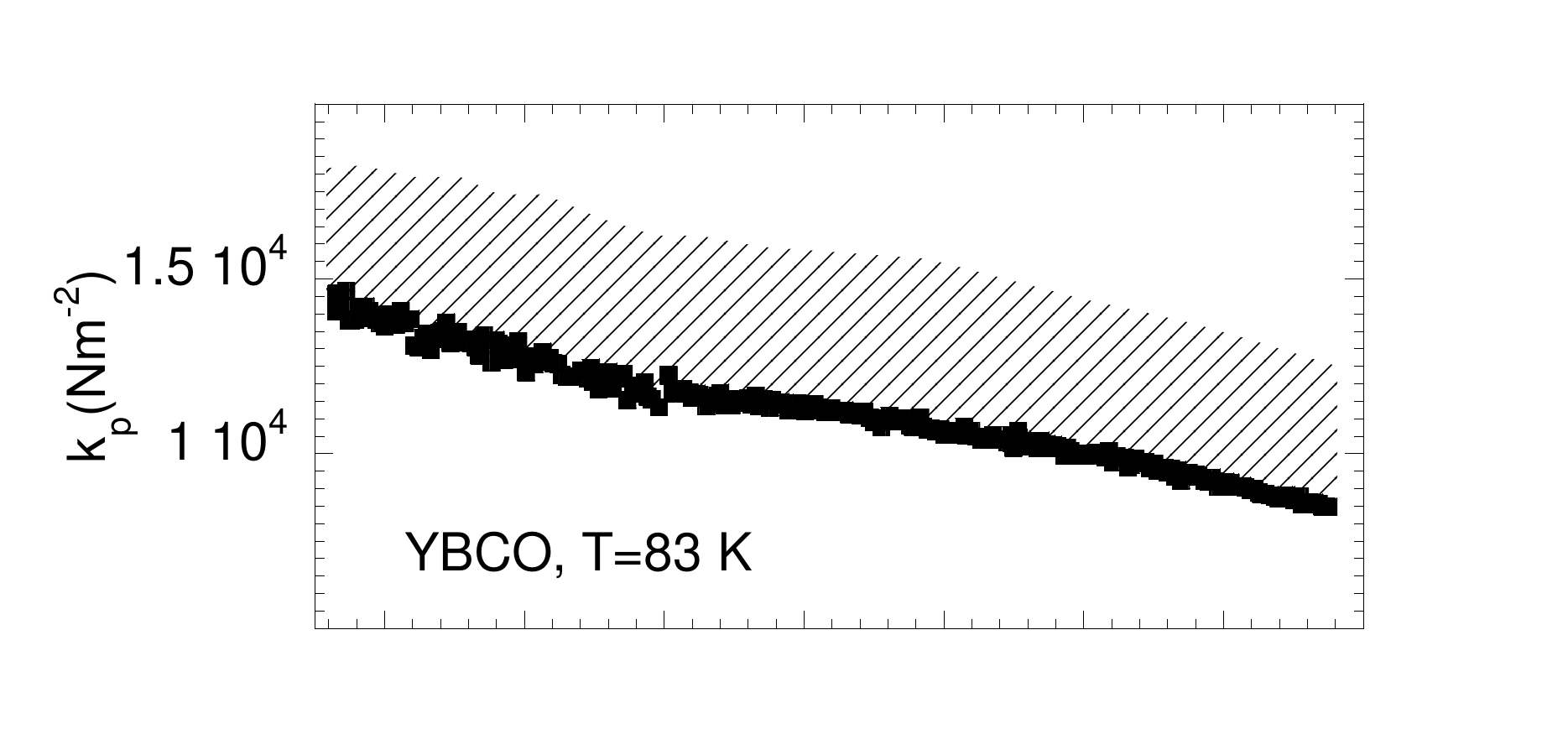}}
\vspace{-11mm}
\centerline{\includegraphics[width=8.3cm]{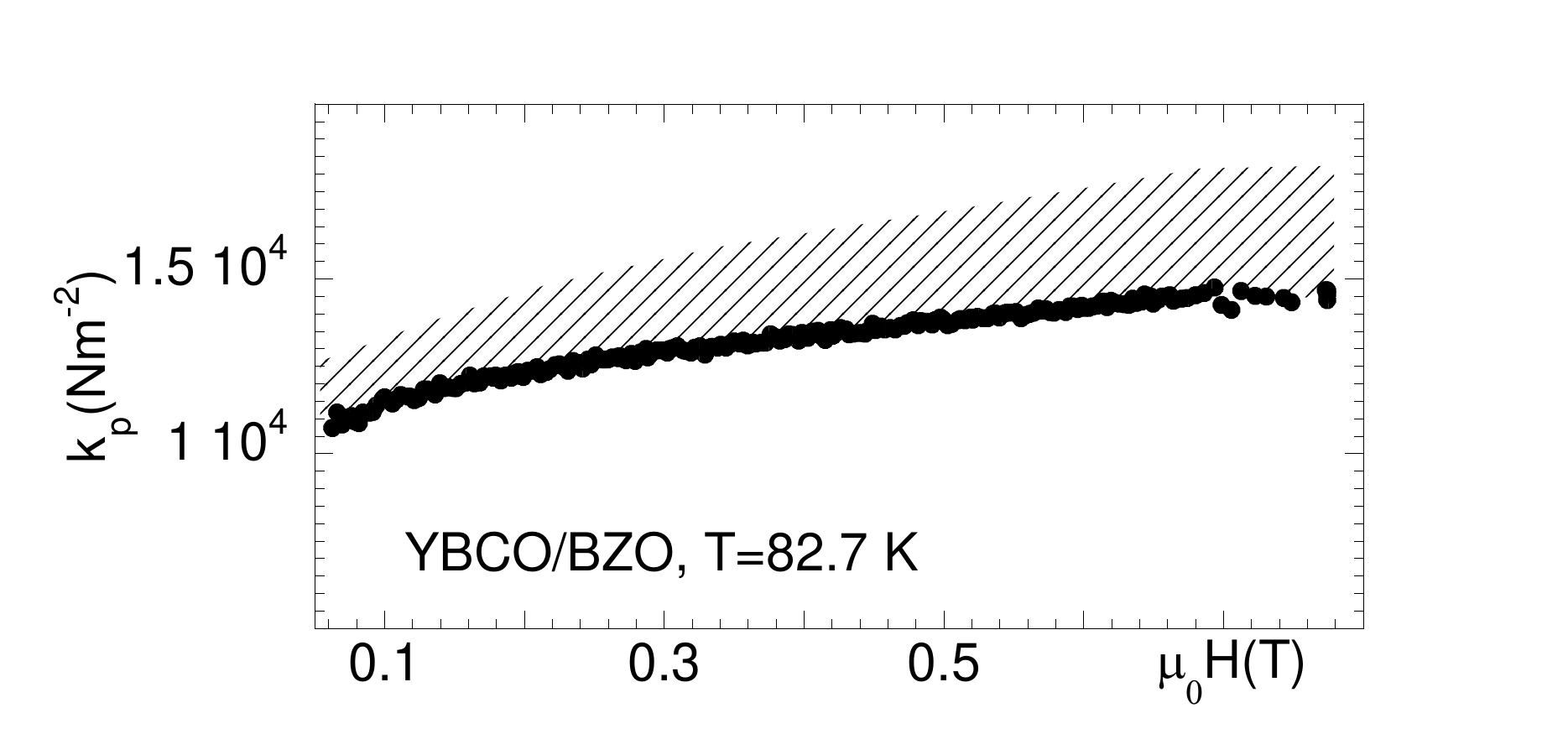}}
\vspace{-4mm}
  \caption{Allowed ranges (shaded areas) and GR limit values (symbols) for $k_p(H)$ (see text) at selected temperatures for pure YBCO and YBCO/BZO. Note the very different field dependence.}
\label{fig_kp}
\end{figure}

Making reference to Eq.(\ref{eqB}), the observed field dependence on $r$ can originate from any of the fluxon parameters involved. While a field varying pinning energy $U(H)$ is a plausible candidate, one cannot a-priori attribute to it the whole field variation of $r(H)$. In order to proceed correctly, one should identify all the relevant field-dependencies. This task is hindered by the under-determination of the system of equations (given by Eq.(\ref{eqB}) ) which relates the unknowns parameters with the experimentally determined $\Delta\rho_1(H)$ and $\Delta\rho_2(H)$. Nevertheless, it is possible to proceed as follows. By making no assumption about the creep parameter $\epsilon'$, one can leave it free to vary within the boundaries already highlighted in the preceding Section (i.e., $0\leq\epsilon'\leq\epsilon'_{max}$). In this way, both $k_p$ and $\eta$ (together to their ratio $\omega_p$) will vary within correspondent bounds which can easily computed from Eq.(\ref{eqB}) with some algebra (full computations will be reported elsewhere \cite{talliotech}). In the following, we focus our attention on $k_p$, given its relevance regarding elastic and pinning properties of the fluxon system. The $k_p(H)$ allowed ranges (shaded areas) and GR limit curves (symbols) are shown in Figure \ref{fig_kp} for both samples for two temperatures; on average, the allowed ranges are $\sim$20$\%$ wide with worst case values reaching $\sim$30$\%$.
Many comments can be done. First of all, the absolute values of $k_p$ in YBCO/BZO are larger than in pure YBCO, especially at lower temperatures (panel `a'), confirming the strong difference in the reactive response already observed. Secondly, an apparent difference in the field dependence can be indeed observed between the two samples: while in pure YBCO $k_p(H)$ steadily decreases, with larger slopes at lower temperatures (lower curve in panel `a', panel `b'), in YBCO/BZO $k_p(H)$ appears to be almost constant (upper curve in panel `a'), or slightly increasing when higher temperatures are considered (panel `c'). These results can find a natural explanation by considering the nature of the pinning sites in the two samples. In pure YBCO, $k_p$ suggests that, among all the available pinning sites, only a small fraction of them is steep enough to be actually effective in pinning vortices, whose oscillations have extremely small amplitudes due to our high measuring frequency. In this scenario, fluxons immediately outnumber the effective pins, so that a fixed overall pinning strength has to be averaged over an increasing number of fluxons, which determines the observed decreasing (mean-field single vortex) $k_p$ \cite{gr,golosovskySUST96}. In YBCO/BZO, on the other hand, BZO inclusions give a high density of strong pins, so that in the whole field range here explored each fluxon entering in the sample can be individually pinned. As a consequence, $k_p$ is almost constant against $H$. These hypotheses can be checked by studying samples with intermediate concentrations of BZO inclusions with respect the samples here presented. In this way, by increasing the artificial pins density, one should expect a progressive reduction of the field dependence of $k_p$ as well as an increase of the absolute values.
\section{Conclusion}
\label{conc}
We have presented data for the microwave resistivity at 47.7 GHz in YBCO films with BZO inclusions and in pure YBCO. The comparison has shown that BZO inclusions strongly reduce the field-induced microwave resistivity. Moreover, its interpretation within the Brandt model revealed that in pure YBCO the drop of elastic response necessarily requires a field dependent pinning constant $k_p$, most likely due to a small density of pinning sites. This is not the case for YBCO/BZO, where the BZO inclusions presumably provides for a high density of strong pins which allows to individually pin each vortex in the whole field regime explored.




\begin{thebibliography}{00}


%

\bibitem{blatterRMP94} G. Blatter et al.,  {Rev. Mod. Phys.} {66} (1994)  1125.
%

\bibitem{civalePRL91} L. Civale et al., {Phys. Rev. Lett.} {67} (1991) 648.

\bibitem{fendrichPRL95}  J. A. Fendrich et al., {Phys. Rev. Lett.} {74} (1995) 1210.

\bibitem{macmanusNATMAT04} J. L. Macmanus-Driscoll et al., {Nat. Mater.} {3} (2004) 439.
%
%
%
%


\bibitem{peurlaPRB07}  M. Peurla et al., {Phys. Rev. B} {75} (2007) 184524.

\bibitem{tomaschPRB88}  W. J. Tomasch et al., {Phys. Rev. B} {37} (1988) 9864.

%
\bibitem{tsuchiyaPRB01} Y. Tsuchiya et al., {Phys. Rev. B} {63} (2001) 184517.

\bibitem{hanaguriPRL99} T. Hanaguri et al.,  {Phys. Rev. Lett.} {82} (1999) 1273.

\bibitem{silvaIJMPB00} E. Silva et al., {Int. J. of Mod. Phys. B} {14} (2000) 2822.

%

\bibitem{gr} J. I. Gittleman, B. Rosenblum, {Phys. Rev. Lett.} {16} (1966) 734.

\bibitem{cc} M. W. Coffey, J. R. Clem, {Phys. Rev. Lett.} {67} (1991) 386.

\bibitem{brandtPRL91} E. H. Brandt, {Phys. Rev. Lett.} {67} (1991) 2219.

\bibitem{kv} A. E. Koshelev, V. M. Vinokur, {Physica C} {173} (1991) 465.

\bibitem{halbritter} J. Halbritter, {J. Supercond.} {8} (1995) 691.

%

\bibitem{galluzziIEEE07} V. Galluzzi et al., {IEEE Trans. Appl. Supercond.} {17} (2007) 3628.

\bibitem{damAPL94} B. Dam et al., {Appl. Phys. Lett.} {65} (1994) 1581.

\bibitem{pompeoJSUP07} N. Pompeo, R. Marcon, E. Silva, {J. Supercond. and Novel Magnetism} {20} (2007) 71.
%
%



\bibitem{pompeoPREP07} N. Pompeo et al., {Supercond. Sci. Technol.} (2007) {20} 1002.

%


\bibitem{talliotech} N. Pompeo, E. Silva, in preparation.

\bibitem{golosovskySUST96} M. Golosovsky, M. Tsindlekht, D. Davidov, {Supercond. Sci. Technol.} {9} (1996) 1.




\end{thebibliography}
\end{document}